\documentclass[journal=jctcce, manuscript=article]{achemso}

\usepackage[table]{xcolor}
\usepackage[version=3]{mhchem}
\usepackage{amsfonts}
\usepackage{amsmath}
\usepackage{amsthm}
\usepackage{amssymb}
\usepackage{graphicx}
\graphicspath{{img/}}
 \DeclareGraphicsExtensions{.pdf} 
\usepackage{caption}
\usepackage{subcaption}
 \DeclareCaptionSubType*[arabic]{figure}
\usepackage{braket}
\usepackage{multicol}
\usepackage{todonotes}
\usepackage{booktabs}
\usepackage[nohyperlinks,nolist]{acronym}
\usepackage{siunitx}
\usepackage{mathtools}
\usepackage{hyperref}
\usepackage[inline]{enumitem}

\newlength{\dch}
\newlength{\sch}

\newcommand{\scaling}{\phi}

\newcommand{\vampyr}{\textit{VAMPyR}}
\newcommand{\grasp}{\textit{GRASP}}
\newcommand{\dirac}{\textit{DIRAC}}

\author{Christian Tantardini}
\affiliation{Hylleraas center, UiT The Arctic University of Norway, PO Box 6050 Langnes, N-9037 Troms\o, Norway}
\alsoaffiliation{Department of Materials Science and NanoEngineering, Rice University, Houston, Texas 77005, United States of America}
\email{christiantantardini@ymail.com}

\author{Roberto Di Remigio Eik\aa s}
\affiliation{Algorithmiq Ltd., Kanavakatu 3C, FI-00160, Helsinki, Finland}
\alsoaffiliation{Hylleraas center, UiT The Arctic University of Norway, PO Box 6050 Langnes, N-9037 Troms\o, Norway}

\author{Magnar Bj{\o}rgve}
\affiliation{Hylleraas center, UiT The Arctic University of Norway, PO Box 6050 Langnes, N-9037 Troms\o, Norway}

\author{Stig Rune Jensen}
\affiliation{Hylleraas center, UiT The Arctic University of Norway, PO Box 6050 Langnes, N-9037 Troms\o, Norway}

\author{Luca Frediani}
\affiliation{Hylleraas center, UiT The Arctic University of Norway, PO Box 6050 Langnes, N-9037 Troms\o, Norway}
\email{luca.frediani@uit.no}

\title{Full Breit Hamiltonian in the Multiwavelets Framework}


\begin{document}

\begin{acronym}
\acro{AO}{atomic orbital}
\acro{API}{Application Programmer Interface}
\acro{AUS}{Advanced User Support}
\acro{BEM}{Boundary Element Method}
\acro{BO}{Born-Oppenheimer}  
\acro{CBS}{complete basis set}
\acro{CC}{Coupled Cluster}
\acro{CTCC}{Centre for Theoretical and Computational Chemistry}
\acro{CoE}{Centre of Excellence}
\acro{DC}{dielectric continuum}  
\acro{DCHF}{Dirac-Coulomb Hartree-Fock}  
\acro{DFT}{density functional theory}  
\acro{DKH}{Douglas-Kroll-Hess}
\acro{EFP}{effective fragment potential}
\acro{ECP}{effective core potential}
\acro{EU}{European Union}
\acro{GGA}{generalized gradient approximation}
\acro{GPE}{Generalized Poisson Equation}
\acro{GTO}{Gaussian Type Orbital}
\acro{HF}{Hartree-Fock}  
\acro{HPC}{high-performance computing}
\acro{Hylleraas}[HC]{Hylleraas Centre for Quantum Molecular Sciences}
\acro{IEF}{Integral Equation Formalism}
\acro{IGLO}{individual gauge for localized orbitals}
\acro{KB}{kinetic balance}
\acro{KS}{Kohn-Sham}
\acro{LAO}{London atomic orbital}
\acro{LAPW}{linearized augmented plane wave}
\acro{LDA}{local density approximation}
\acro{MAD}{mean absolute deviation}
\acro{maxAD}{maximum absolute deviation}
\acro{MM}{molecular mechanics}  
\acro{MCSCF}{multiconfiguration self consistent field}
\acro{MPA}{multiphoton absorption}
\acro{MRA}{multiresolution analysis}
\acro{MSDD}{Minnesota Solvent Descriptor Database}
\acro{MW}{multiwavelet}
\acro{NAO}{numerical atomic orbital}
\acro{NeIC}{nordic e-infrastructure collaboration}
\acro{KAIN}{Krylov-accelerated inexact Newton}
\acro{NMR}{nuclear magnetic resonance}
\acro{NP}{nanoparticle}  
\acro{NS}{non-standard}  
\acro{OLED}{organic light emitting diode}
\acro{PAW}{projector augmented wave}
\acro{PBC}{Periodic Boundary Condition}
\acro{PCM}{polarizable continuum model}
\acro{PW}{plane wave}
\acro{QC}{quantum chemistry}  
\acro{QM/MM}{quantum mechanics/molecular mechanics}  
\acro{QM}{quantum mechanics}  
\acro{RCN}{Research Council of Norway}
\acro{RMSD}{root mean square deviation}
\acro{RKB}{restricted kinetic balance}
\acro{SC}{semiconductor}
\acro{SCF}{self-consistent field}
\acro{STSM}{short-term scientific mission}
\acro{SAPT}{symmetry-adapted perturbation theory}
\acro{SERS}{surface-enhanced raman scattering}
\acro{WPREL}[WP1]{Work Package 1}
\acro{WPROP}[WP2]{Work Package 2}
\acro{WPAPP}[WP3]{Work Package 3}
\acro{WP}{Work Package}  
\acro{X2C}{exact two-component}
\acro{ZORA}{zero-order relativistic approximation}
\acro{ae}{almost everywhere}
\acro{BVP}{boundary value problem}
\acro{PDE}{partial differential equation}
\acro{RDM}{1-body reduced density matrix}
\acro{SCRF}{self-consistent reaction field}
\acro{IEFPCM}{Integral Equation Formalism \ac{PCM}}
\acro{FMM}{fast multipole method}
\acro{DD}{domain decomposition}
\acro{TRS}{time-reversal symmetry}
\acro{SI}{Supporting Information}
\acro{DHF}{Dirac--Hartree--Fock}
\end{acronym}



\begin{abstract}
New techniques in core-electron spectroscopy are necessary to resolve the structures of oxides of \textit{f}-elements and other strongly correlated materials that are present only as powders and not as single crystals.
Thus, accurate quantum chemical methods need to be developed to calculate core spectroscopic properties in such materials.
In this contribution, we present an important  development in this direction, extending our fully adaptive real-space multiwavelet basis framework to tackle the 4-component Dirac-Coulomb-Breit Hamiltonian.
We show that Multiwavelets are able to reproduce one-dimensional grid-based approaches. They are however a fully three-dimensional approach which can later on be extended to molecules and materials.
Our Multiwavelet implementation attained precise results irrespective of the chosen nuclear model, provided that the error threshold is tight enough and the chosen polynomial basis is sufficiently large.
Furthermore, our results confirmed that in two-electron species, the magnetic and Gauge contributions from \textit{s}-orbitals are identical in magnitude and can account for the experimental evidence from $K$ and $L$ edges.
\end{abstract}

\begin{center}
    \includegraphics[width=0.5\textwidth]{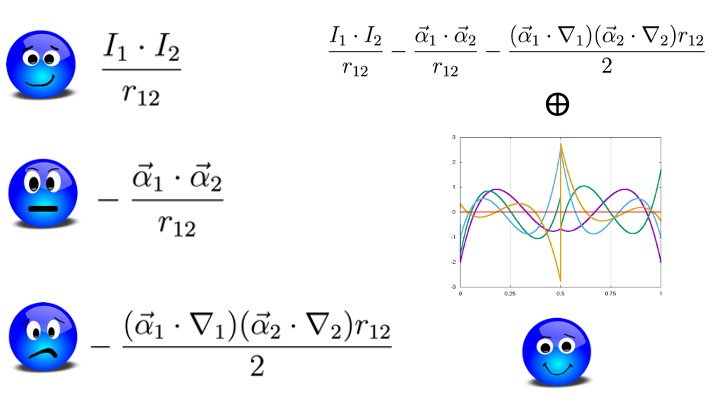}
\end{center}

\section{Introduction}
Core-electron spectroscopies like X-ray photoelectron spectroscopy, X-ray absorption spectroscopy and electron energy loss spectroscopy are powerful tools to investigate the electronic structure of transition-metal and rare-earth materials.
For example, multi-layered transition metal carbides and carbonitrides \ce{M_{n+1}AX_{n}}, where \ce{M} is an early transition metal, \ce{A} is an A-group element (mostly groups 13 and 14), \ce{X} is \ce{C} or/and \ce{N} and $n$ is 1 to 3 \cite{Naguib_2011}.
These materials can be employed for
energy storage systems, such as lithium-ion batteries \cite{Naguib_2011,Naguib_2012,Naguib_2013,Mashtalir_2013,Tang_2012},
lithium-ion capacitors \cite{Come_2012}, aqueous pseudocapacitors \cite{Ghidiu_2014,Lukatskaya_2013},
and transparent conductive films \cite{Halim_2014}.
Additionally, rare earths are contained in transparent conducting oxides which are considered the new frontier in the area of optoelectronics\cite{Chandiramouli_2013,Liu_2018,Dixon_2016}.
These materials have the unique behaviour of being both optically transparent and electrically conducting which makes them key components
in many optoelectronic devices such as solar cells, flat panel
displays, thin film transistors, and light emitting diodes \cite{Chandiramouli_2013,Liu_2018,Dixon_2016}.

Unfortunately, their spectra are not straightforwardly interpretable due to relativistic effects.
All relativistic effects such as spin-orbit interactions, electron-electron interaction in the valence shell, and between core and valence electrons, will play a role in the core electron spectra\cite{Grobe_1993,Lundqvist_1969,Wendin_1976,Brus_1984,Pines_1956,Gusev_1995,Mulazzi_2010,Lee_2000,Kahk_2014,Glatzel_2005}.
A computational approach based on first-principle calculations that will take into account both relativity and electron correlation could help the interpretation of such spectra.
A recent, promising approach in quantum chemistry is based on multiresolution analysis (MRA), by making use of \acp{MW} \cite{Harrison_2003}. This method has gained momentum in recent years, and has been applied to compute \ac{CBS} limit results for energies and linear response properties of a large number of compounds both within \ac{HF} and \ac{DFT}\cite{Yanai_2005,Vence_2012,Yanai_2015,Jensen2016-vb,Brakestad_polarizabilities}.
A variational treatment of relativistic effects into MRA will allow modelling the spectra of transition metal and rare earth materials. An important step in this direction was presented to tackle the mean-field atomic and molecular Dirac-Coulomb problem in an adaptive, 4-component multiwavelets basis \cite{Dirac_1928, Anderson_2019}.
In such a model the electrons are considered static charges where the average interaction between electrons is modelled with the Coulomb-like term only.
This is the lowest-order relativistic approximation for the two-electron interaction, which disregards the magnetic interactions, such as spin-other-orbit, and the retardation effects due to the finite speed of light.
These effects are important and must be taken into account for a realistic modelling of core-electron spectroscopies. Therefore,  \cite{Mussard_2018,Petrov_2004,Vidal_2020,Kasper_2020}
the Breit interaction terms must thus be included \cite{Breit_1928,Breit_1932,Moss_2012,Dyall_2007,Helgaker_2012}.
The Breit Hamiltonian adds two negative terms, called Gaunt and Gauge, respectively:

\begin{equation}\label{eq:breit_ham}
\begin{split}
     \hat{H}^{Coulomb} + \hat{H}^{Breit} &= \hat{H}^{Coulomb} + \hat{H}^{Gaunt} + \hat{H}^{Gauge} \\
     &= 
     \frac{I_{1}\cdot I_{2}}{r_{12}} 
     - \frac{\Vec{\alpha}_{1} \cdot \Vec{\alpha}_{2}}{r_{12}}
     - \frac{(\Vec{\alpha}_{1} \cdot \nabla_{1}) (\Vec{\alpha}_{2} \cdot \nabla_{2})r_{12}}{2}
\end{split}
\end{equation}

The first term in Eq.~\eqref{eq:breit_ham} is the non-relativistic Coulomb interaction.
The second term, called Gaunt, can be seen, in the non-relativistic limit, as the scalar product between the curl of two spin orbitals: $\Vec{\alpha}_{i} \sim \nabla \times \phi_{i}$ \cite{Moss_2012}.
$\Vec{\alpha}$ denotes a Cartesian vector collecting the $4\times 4$ Dirac
matrices $\alpha_{x}$, $\alpha_{y}$ and $\alpha_{z}$ (\emph{vide infra}).   
When $\Vec{\alpha}$ acts on a 4-component orbital, it couples its
components, as detailed later on in this contribution.
This means that the spin rotation of one electron on its axis generates a
vector potential that will interact with the vector potentials generated by all
other electrons present in the system,\cite{Moss_2012} resulting in a scalar potential.
Finally, the third term, called Gauge, describes the retardation effects due to
the reciprocal interaction between the rotational vector fields  $(\alpha_{i}
\cdot \nabla_{i})$ of two electrons  \cite{Moss_2012}.
These contributions cannot be neglected in systems that contain heavy or
super-heavy elements, especially in the calculation of core spectroscopic
properties \cite{Mussard_2018,Petrov_2004,Vidal_2020,Kasper_2020}.

In this contribution, we will present the adaptive MRA multiwavelet
implementation of the \emph{full} Breit interaction as a perturbative
correction on top of a 4-component Dirac-Coulomb-Hartree-Fock wavefunction.
We will demonstrate the precision of our implementation by comparing
ground-state energies of highly-charged helium-like ions with increasing $Z$,
\ce{X^{(Z-2)+}},  performed with our Python code, \vampyr{} (Very Accurate Multiresolution
Python Routines) \cite{battistella_2021} with numerical
radial integration in \grasp{}\cite{Jonsson2013-co} and Gaussian basis
set calculations with the \dirac{}\cite{Saue2020-ce} software.

\section{Theory and Implementation}

\subsection{Multiresolution Analysis and Multiwavelets}

Multiresolution analysis\cite{Alpert_MRA} is constructed by considering a set
of orthonormal functions called \emph{scaling} functions $\scaling_i(x)$
supported on the interval $[0,1]$. They can be dilated and translated to obtain
a corresponding basis in subintervals of $[0,1]$. The most common procedure is
a dyadic subdivision, such that at scale $n$ there will be $2^n$ intervals
defined by a translation index $l=0,2^n-1$ such that the scaling functions in
the $l$-th interval $[l/2^n, (l+1)/2^n]$ are obtained as:
\begin{equation}
    \scaling^n_{il} = 2^{n/2} \scaling_i(2^n x - l)
\end{equation}
Additionally, functions at subsequent scales are connected by the
\emph{two-scale relationships} which allow to obtain the scaling function at
scale $n$ as a linear combination of scaling functions at scale $n-1$. 

This construction leads to a ladder of scaling spaces in a telescopic sequence
which is dense in $L^2$:
\begin{equation}
 V_{0}^{k} \subset V_{1}^{k} \subset ..... V_{n}^{k}\subset ....\subset L^{2}
\end{equation}

The \emph{multiwavelet functions} are then obtained as the orthogonal
complement of the scaling functions at scale $n+1$ with respect to the ones at
scale $n$.

\begin{equation}
 V_{n}^{k} \oplus W_{n}^{k} = V_{n+1}^{k},\: \: \: W_{n}^{k} \perp V_{n}^{k}
\end{equation}

In the construction of Alpert\cite{Alpert_1993}, the scaling functions are a
simple set of polynomials, and the wavelet functions are then piecewise
polynomial functions. The possibility to construct efficient algorithms, with
precise error control, relies on the combination of several properties of such
a construction. Here, it will suffice to say that the most important aspects
concern the disjoint support of the basis, which enables function-based
adaptivity, the vanishing moments of the wavelet functions, which guarantees
fast decay of the representation coefficients, the \ac{NS} form of
operators\cite{Beylkin_ACHA}, which uncouples scales during operator
application thus preserving adaptivity, the separated representation of
integral kernels, which leads to low-scaling algorithms. The interested reader
is referred to the available literature for details about those
aspects.\cite{Alpert_1993, Beylkin2005-kg, Frediani_2013, Harrison_2003}

\subsection{Mean-field two-electron operators in a Multiwavelets Basis}

We will summarise the main methodological developments enabling the results in
this contribution. We first recall that in a relativistic framework, molecular
orbitals are vectors with four complex components. We will use indices:
\begin{itemize}
\item $u, w \in \lbrace x, y, z \rbrace$ for Cartesian components, 
\item $p, q, \ldots$ for occupied 4-component orbitals,
\item $A, B, \ldots \in \lbrace 1, 2, 3, 4 \rbrace$ for orbital components.
\end{itemize}
Furthermore, Greek capital letters will be used for the 4-component orbitals
and their lowercase counterparts will be used for the corresponding components:
\begin{equation}
    \Phi_{p}
    = 
    \begin{pmatrix}
    \varphi_{p}^{1} \\
    \varphi_{p}^{2} \\
    \varphi_{p}^{3} \\
    \varphi_{p}^{4}
    \end{pmatrix}
\end{equation}
The corresponding Hermitian conjugate (transposed and complex conjugate)
orbital is:
\begin{equation}
    \Phi_{p}^{\dagger}
    = 
    \begin{pmatrix}
    \overline{\varphi}_{p}^{1} & 
    \overline{\varphi}_{p}^{2} &
    \overline{\varphi}_{p}^{3} &
    \overline{\varphi}_{p}^{4}
    \end{pmatrix}
\end{equation}
with $\dagger$ denoting Hermitian conjugation and overline complex conjugation
of a component.

To avoid confusion we will also refer to the instantaneous electron interaction (first term in Eq.~\ref{eq:breit_ham} as the \emph{Coulomb} term, whereas we will use the terms \emph{direct} and \emph{exchange} to refer to the two parts of each term, arising from the fermionic nature of the electrons.

For the Coulomb operator $g^{Coulomb}(\Vec{r}_1,\Vec{r}_2) = \frac{I_1 \cdot I_2}{r_{12}}$, the direct and exchange operators are straightforward and shown in Eqs.~(7a) and (7b) in the Supporting Information, respectively. 
In practice, these operators are applied as convolutions. Efficient and accurate convolution with an integral operator is implemented in a separated representation (see Ref.~\citenum{Frediani_2013} for details).
We underline that the Coulomb part of the two-electron interaction is in this
framework \emph{diagonal}, in the sense that it is not coupling the four
components of the spinor. In a \ac{GTO} framework, the exchange part would
instead couple the four components of the spinor, because the formalism is tied
to the \ac{AO} densities, thus generating an artificial coupling once the
exchange operation is performed\cite{Reiher2014-cp}. 

We proceed similarly for the Gaunt operator $g^{Gaunt}(\Vec{r}_1,\Vec{r}_2) = - \frac{\Vec{\alpha}_1 \cdot \Vec{\alpha}_2}{r_{12}}$.
Note that the $\Vec{\alpha}$ appearing in the numerator are Cartesian vectors
whose components are $4\times 4$ anti-diagonal block matrices:
\begin{equation}
\alpha_{u} = 
\begin{pmatrix}
0 & \sigma_{u} \\
\sigma_{u} & 0
\end{pmatrix}
\end{equation}
with $\sigma_{u},\, u \in \lbrace x, y, z\rbrace$, the Pauli matrices. Applying
$\alpha_{u}$ on a 4-component orbital, in practice reorders the components,
possibly multiplied by a phase factor.

The two-electron energy for the Gaunt operator is thus:
\begin{align}
   E^{Gaunt}
    =
    & -
   \frac{1}{2}
   \sum_{pq}
   \int d \Vec{r}_1\int d \Vec{r}_2
   \frac{\Vec{j}_{pp}(\Vec{r}_1)\cdot \Vec{j}_{qq}(\Vec{r}_2) }{r_{12}} \\
   & +
   \frac{1}{2}
   \sum_{pq}
   \int d \Vec{r}_1\int d \Vec{r}_2
   \frac{\Vec{j}_{pq}(\Vec{r}_1)\cdot \Vec{j}_{qp}(\Vec{r}_2) }{r_{12}}
\end{align}
where we have introduced the current density Cartesian vector, with components:
\begin{equation}
   j_{pq;u} = \sum_{AB} \overline{\varphi}^{A}_{p}\alpha^{AB}_{u}\varphi^{B}_{q},
\end{equation}
to rewrite the expression more compactly. 
The corresponding mean-field, effective one-electron, direct and exchange operators are:
\begin{subequations}
 \begin{align}
J^{Gaunt}\Phi_{k}
&= 
\sum_{u}
\left[
\int d \Vec{r}_2
\frac{\sum_{q} \Phi^{\dagger}_{q}(\Vec{r}_2)\alpha_{u}\Phi_{q}(\Vec{r}_2)}{|\Vec{r}_1 - \Vec{r}_2|}
\right]
\alpha_{u}
\Phi_{k}(\Vec{r}_1) 
=
\left[
\int d \Vec{r}_2
\frac{\Vec{j}(\Vec{r}_2)}{|\Vec{r}_1 - \Vec{r}_2|}
\right]
\cdot
\left[
\Vec{\alpha}
\Phi_{k}
\right]
\label{eq:Gaunt-direct} \\
K^{Gaunt}\Phi_{k} 
&= 
\sum_{q} 
\sum_{u}
\alpha_u \Phi_{q}(\Vec{r}_1) 
\left[
\int d \Vec{r}_2
\frac{j_{qk;u}(\Vec{r}_2)}{|\Vec{r}_1 - \Vec{r}_2|}
\right]
=
\sum_{q} \left[ \Vec{\alpha}\Phi_{q}\right] \cdot \Vec{V}^{Gaunt}_{qk}
,
\label{eq:Gaunt-exchange}
 \end{align}
\end{subequations}
$\Vec{j}$ is the trace of the matrix collecting the orbital-pair current densities $j_{pq;u}$.

The Gaunt direct and exchange operators use the same primitive as the Coulomb
operators for the convolution with the inverse-distance kernel. Thus:
\begin{enumerate}
    \item Although the expressions for the Gaunt mean-field operators appear more complicated than those stemming from the Coulomb interaction, their computational load is only three times higher, because each component of the $\Vec{\alpha}$ vector only has four non-zero elements.
    \item For each Cartesian component, one can compute a ``Gaunt potential" which is then multiplied by the $\Vec{\alpha}$-transformed orbital, exactly as for the Coulomb interaction.
\end{enumerate}

Turning our attention to the gauge two-electron potential, we follow the suggestion of Sun \emph{et al.}\cite{Sun_2022}
$
    -\nabla_{1} \frac{1}{r_{12}}
    \equiv
    \frac{\Vec{r}_{12}}{r_{12}^3}
    \equiv
    \nabla_{2} \frac{1}{r_{12}},
$
and rewrite it as:
\begin{align}
g^{Gauge}(\Vec{r}_1,\Vec{r}_2) 
&= 
\frac{1}{2} \frac{(\Vec{\alpha}_1 \cdot \Vec{r}_{12})(\Vec{\alpha}_2 \cdot \Vec{r}_{12})}{r_{12}^3} \\
&=\label{eq:gauges}
    -\frac{1}{2}
\left[\Vec{\alpha}_1 \cdot \left( \mp \nabla_{1,2} \frac{1}{r_{12}}\right)\right]
\left(\Vec{\alpha}_2 \cdot \Vec{r}_{1} \right)
+
\frac{1}{2}
\left[\Vec{\alpha}_1 \cdot \left( \mp\nabla_{1,2} \frac{1}{r_{12}}\right)\right]
\left(\Vec{\alpha}_2 \cdot \Vec{r}_{2}\right)
\end{align}
where the sign/index pairs ($-\nabla_{1}$ or $+\nabla_{2}$) can be chosen independently for each
of the two terms, giving rise to \emph{four} equivalent expression.

The energy expressions corresponding to each of the above forms can be
considerably simplified using integration by parts, thus avoiding the need for
differentiating the inverse-distance kernel. However, of the four forms
presented above, the energy expression obtained by choosing $+\nabla_2$ in both
terms of Eq.~\eqref{eq:gauges} is the most compact \emph{and} computationally
parsimonious:
\begin{align}
 E^{Gauge}
   & =
   \frac{1}{2}
   \sum_{pq}
   \int d \Vec{r}_1\int d \Vec{r}_2
   \frac{\left(\Vec{j}_{pp}(\Vec{r}_1)\cdot \Vec{r}_{1}\right) \left(\nabla_{2}\cdot\Vec{j}_{qq}(\Vec{r}_2)\right)}{2r_{12}} \label{eq:gauge-4-energy-1term} \\
   & - 
   \frac{1}{2}
   \sum_{pq}
   \int d \Vec{r}_1\int d \Vec{r}_2
   \frac{\left(\Vec{j}_{pp}(\Vec{r}_1)\cdot \Vec{r}_{2}\right) \left(\nabla_{2}\cdot\Vec{j}_{qq}(\Vec{r}_2)\right)}{2r_{12}} \label{eq:gauge-4-energy-2term} \\
   & - 
   \frac{1}{2}
   \sum_{pq}
   \int d \Vec{r}_1\int d \Vec{r}_2
   \frac{\Vec{j}_{pp}(\Vec{r}_1)\cdot \Vec{j}_{qq}(\Vec{r}_2)}{2r_{12}} \label{eq:gauge-4-energy-3term} \\
   & -
   \frac{1}{2}
   \sum_{pq}
   \int d \Vec{r}_1\int d \Vec{r}_2
   \frac{\left(\Vec{j}_{pq}(\Vec{r}_1)\cdot \Vec{r}_{1}\right) \left(\nabla_{2}\cdot\Vec{j}_{qp}(\Vec{r}_2)\right)}{2r_{12}} \label{eq:gauge-4-energy-4term} \\
   & + 
   \frac{1}{2}
   \sum_{pq}
   \int d \Vec{r}_1\int d \Vec{r}_2
   \frac{\left(\Vec{j}_{pq}(\Vec{r}_1)\cdot \Vec{r}_{2}\right) \left(\nabla_{2}\cdot\Vec{j}_{qp}(\Vec{r}_2)\right)}{2r_{12}} \label{eq:gauge-4-energy-5term} \\
   & + 
   \frac{1}{2}
   \sum_{pq}
   \int d \Vec{r}_1\int d \Vec{r}_2
   \frac{\Vec{j}_{pq}(\Vec{r}_1)\cdot \Vec{j}_{qp}(\Vec{r}_2)}{2r_{12}} \label{eq:gauge-4-energy-6term}
\end{align}

The former three terms are the direct contributions and the latter three
the exchange contributions. The use of the inverse-distance kernel is the most
significant advantage of this formulation, since that is already an efficient
and robust computational primitive in a multiwavelet basis. Note that the
calculation of the divergence of the orbital current densities 
$$\nabla\cdot\Vec{j}_{pq} \equiv 
\frac{\partial j_{pq;x}}{\partial {x}} +
\frac{\partial j_{pq;y}}{\partial {y}} +
\frac{\partial j_{pq;z}}{\partial {z}}$$ 
is both efficient and precise in a multiwavelet basis.\cite{Anderson2019-bx}

Finally, we present the expressions for the direct and exchange Gauge
mean-field operators:
\begin{subequations}
\begin{align}
J^{Gauge}\Phi_{k}
& =
\begin{aligned}[t]
&\frac{1}{2} \Bigg\lbrace
\left[
\int d \Vec{r}_2
\frac{\nabla_{2}\cdot \Vec{j}(\Vec{r}_2)}{|\Vec{r}_1 - \Vec{r}_2|}
\right]
\left[
\left(
\Vec{\alpha}
\Phi_{k}
\right)
\cdot
\Vec{r}_1
\right]
-
\left[
\int d \Vec{r}_2
\frac{\Vec{j}(\Vec{r}_2)}{|\Vec{r}_1 - \Vec{r}_2|}
\right]  \cdot
\left[
\Vec{\alpha}
\Phi_{k}
\right]  \\
 & -
\left[
\int d \Vec{r}_2
\frac{\Vec{r}_2 \left(\nabla_{2}\cdot \Vec{j}(\Vec{r}_2)\right)}{|\Vec{r}_1 - \Vec{r}_2|}
\right]
\cdot
\left[
\Vec{\alpha}
\Phi_{k}
\right]
\Bigg\rbrace
\end{aligned}
\label{eq:Gauge-direct}
\\
K^{Gauge}\Phi_{k} 
& = 
\begin{aligned}[t]
&\frac{1}{2} \sum_{q} \Bigg\lbrace
\left[
\int d \Vec{r}_2
\frac{\nabla_{2}\cdot \Vec{j}_{qk}(\Vec{r}_2)}{|\Vec{r}_1 - \Vec{r}_2|}
\right]
\left[
\left(
\Vec{\alpha}
\Phi_{k}
\right)
\cdot
\Vec{r}_1
\right]
-
\left[ \Vec{\alpha}\Phi_{q}\right] \cdot \Vec{V}^{Gaunt}_{qk}  \\
& -
\left[
\int d \Vec{r}_2
\frac{\Vec{r}_2 \left(\nabla_{2}\cdot \Vec{j}_{qk}(\Vec{r}_2)\right)}{|\Vec{r}_1 - \Vec{r}_2|}
\right]
\cdot
\left[
\Vec{\alpha}
\Phi_{k}
\right]
\Bigg\rbrace.
\end{aligned}
\label{eq:Gauge-exchange}
\end{align}
\end{subequations}
All terms in both the direct and exchange operators are applied using the
inverse-distance integral operator only.

For completeness, we report also the expressions for the Gauge term when
using the inverse-cube-distance form for the operator:
\begin{equation}
g^{Gauge}(\Vec{r}_1,\Vec{r}_2) 
= 
-
\frac{\left(\Vec{\alpha}_1 \cdot \Vec{r}_{12}\right) \left( \Vec{\alpha}_2 \cdot \Vec{r}_{12}\right)}{2r_{12}^3},
\end{equation}
The two-electron energy reads:
\begin{align}
E^{Gauge}
   =
   & -\frac{1}{2}
   \sum_{pq}
   \int d \Vec{r}_1\int d \Vec{r}_2
   \frac{\left(\Vec{j}_{pp}(\Vec{r}_1)\cdot \Vec{r}_{12}\right) \left(\Vec{j}_{qq}(\Vec{r}_2)\cdot \Vec{r_{12}}\right)}{r_{12}^3} \\
   & +
   \frac{1}{2}
   \sum_{pq}
   \int d \Vec{r}_1\int d \Vec{r}_2
   \frac{\left(\Vec{j}_{pq}(\Vec{r}_1)\cdot \Vec{r}_{12}\right) \left(\Vec{j}_{qp}(\Vec{r}_2)\cdot \Vec{r_{12}}\right)}{r_{12}^3} \label{eq:ga}.
\end{align}

While this is arguably more compact than the sum of all six terms in the previous equation (Eqs.~\eqref{eq:gauge-4-energy-1term}-~\eqref{eq:gauge-4-energy-6term}),  it has
two main disadvantages. First, it is harder to appreciate the physical content
of the expression at a glance. Second, it requires the application of a
different convolution operator.
The latter point is apparent when looking at the expressions for the direct and
exchange operators:

\begin{subequations}
 \begin{align}
J^{Gauge}\Phi_{k}
& = 
\sum_{uw}
\left[
\int d \Vec{r}_2
\frac{(\Vec{r}_1 - \Vec{r}_2)_u (\Vec{r}_1 - \Vec{r}_2)_w}{|\Vec{r}_1 - \Vec{r}_2|^3}
j_{w}(\Vec{r}_2)
\right]
\alpha_{u}
\Phi_{k} \nonumber \\
& =
\left[
\int d \Vec{r}_2
\mathbb{G}(\Vec{r}_1, \Vec{r}_2) \Vec{j}(\Vec{r}_{2})
\right]
\cdot
\Vec{\alpha}
\Phi_{k}
\label{eq:gauge-direct} \\
K^{Gauge}\Phi_{k} 
& = 
\sum_{q} 
\sum_{uw}
\alpha_u \Phi_{q}
\left[
\int d \Vec{r}_2
\frac{(\Vec{r}_1 - \Vec{r}_2)_u (\Vec{r}_1 - \Vec{r}_2)_w}{|\Vec{r}_1 - \Vec{r}_2|^3}
j_{qk;w}(\Vec{r}_2)
\right] \nonumber \\
& =
\sum_{q} \left[ \Vec{\alpha}\Phi_{q}\right] \cdot 
\left[
\int d \Vec{r}_2
\mathbb{G}(\Vec{r}_1, \Vec{r}_2) \Vec{j}_{qk}(\Vec{r}_{2})
\right]
,
\label{eq:gauge-exchange}
 \end{align}
\end{subequations}
The new convolution operator, $\mathbb{G}$, is a
\emph{matrix} convolution operator with 6 unique elements, each of which must be
implemented by approximating the integral representation of the
inverse-cube-distance kernel\cite{Shiozaki_2013} as a finite exponential
sum:\cite{Hackbusch2006-ys}
\begin{equation}
   \frac{(\Vec{r}_1 - \Vec{r}_2)_u (\Vec{r}_1 - \Vec{r}_2)_w}{|\Vec{r}_1 - \Vec{r}_2|^3}
    \simeq
    \sum_{\kappa}a_{\kappa}
    (\Vec{r}_1 - \Vec{r}_2)_u (\Vec{r}_1 - \Vec{r}_2)_w
    \exp(-b_{\kappa}|\Vec{r}_1 - \Vec{r}_2|^2).
\end{equation}
Each term, though anisotropic, can be applied in each Cartesian direction
separately. Coefficients and exponents in the sum are obtained similarly to
those for the inverse-distance convolution operator, see
Ref.~\citenum{Frediani_2013} for details.
This form has been tested in our code, but it turned out to be less stable numerically and significantly more demanding computationally.

\section{Computational Details}
\dirac{} calculations were performed using a nuclear point-charge model and a threshold of $10^{-7}$ on the norm of the error vector (electronic
gradient) was chosen as the convergence criterion for the SCF procedure.
The chosen basis set for \ce{He}, \ce{Ne^{8+}}, \ce{Ar^{16+}},
\ce{Kr^{34+}}, \ce{Xe^{52+}} and \ce{Rn^{84+}} was dyall-aug-cvqz
\cite{Dyall_1998,Dyall_2002,Dyall_2007}.
Furthermore, the calculations were performed using default settings for 4-center integral screening and replacing $(SS|SS)$ integrals by a simple Coulombic correction.
In our MW implementation it is not possible to perform such a correction, because 4-center integrals do not appear in the formalism.
We investigated whether this could impact our perturbative/variational comparisons: with the full two-electron integral tensors the total energy computed with \dirac{} changes slightly and computational cost increases \emph{significantly}. 
However, the \emph{relative error} with respect to both our implementation in \textit{VAMPyR} and in \textit{GRASP} was practically unaffected. This shows that the error is dominated by the intrinsic limitation of the basis set.

\section{Results and Discussion}

We present results for closed-shell, helium-like species: the core $1s$-orbitals are doubly occupied and our code explicitly enforces Kramers' \ac{TRS} \cite{Kramers_1930,Wigner_1932}, such that the 4-component
$1s^{\alpha}$ is related to $1s^{\beta}$ by a quaternionic unitary transformation \cite{Saue_1996}. 

In a mean-field treatment -- \emph{e.g.} \ac{HF} and Kohn-Sham \ac{DFT} -- the Coulomb two-electron operator is replaced by the corresponding \textit{Direct} and \textit{Exchange} terms, indicated with $J$ and $K$, respectively.
Further inclusion of the Gaunt and Gauge interactions in Eq.~\eqref{eq:breit_ham} will
result into additional $J$- and $K$-like terms.
Making use of Kramers \ac{TRS} has a significant impact on the computational cost: the Coulomb interaction will only encompass the \textit{direct} term, whereas \textit{exchange} one will be equal to zero. The Gaunt and Gauge interactions will give rise to both \textit{direct} and \textit{exchange} terms but several contributions will either vanish or be identical to each other.

Previous work by Anderson \textit{et al.}\cite{Anderson_2019} on \emph{full}
4-component Dirac-Coulomb relativistic calculations used smeared nuclear charge
models \cite{Visscher_1997}.
In particular for the isolated atoms they used the Fermi nuclear model
\cite{Visscher_1997}.
This was done to mitigate numerical issues treating core orbitals with a
point-charge model and improve precision.
The Fermi model represents the nuclear charge using
the Fermi-Dirac distribution for the nuclear charge density, introducing two
parameters: the skin thickness and the half-charge radius.
The former is set to 2.30 fm (2.30$\times 10^{-5}$~\AA) for all nuclei
\cite{Visscher_1997}.
The latter is the radius of a sphere containing half of the total nuclear
charge. This parameter depends on the atomic mass of the nucleus $M_N$, with one
expression used when $M_N \leq 5$ atomic mass units and another for $M_N > 5$.\cite{Visscher_1997}
The Fermi model for the nuclear charge is smooth and is thus more physically
meaningful. Furthermore, it avoids singularities at the nuclei, in
contrast to a point-like model.
However, the results of Anderson \textit{et al.} \cite{Anderson_2019} showed
that the achieved precision of multiwavelet methods with respect to the
grid-based approach available in \grasp{} decreases with increasing $Z$,
even though a more physically motivated nuclear model was used.

\begin{figure}[!ht]
    \centering
    \includegraphics[width=\textwidth]{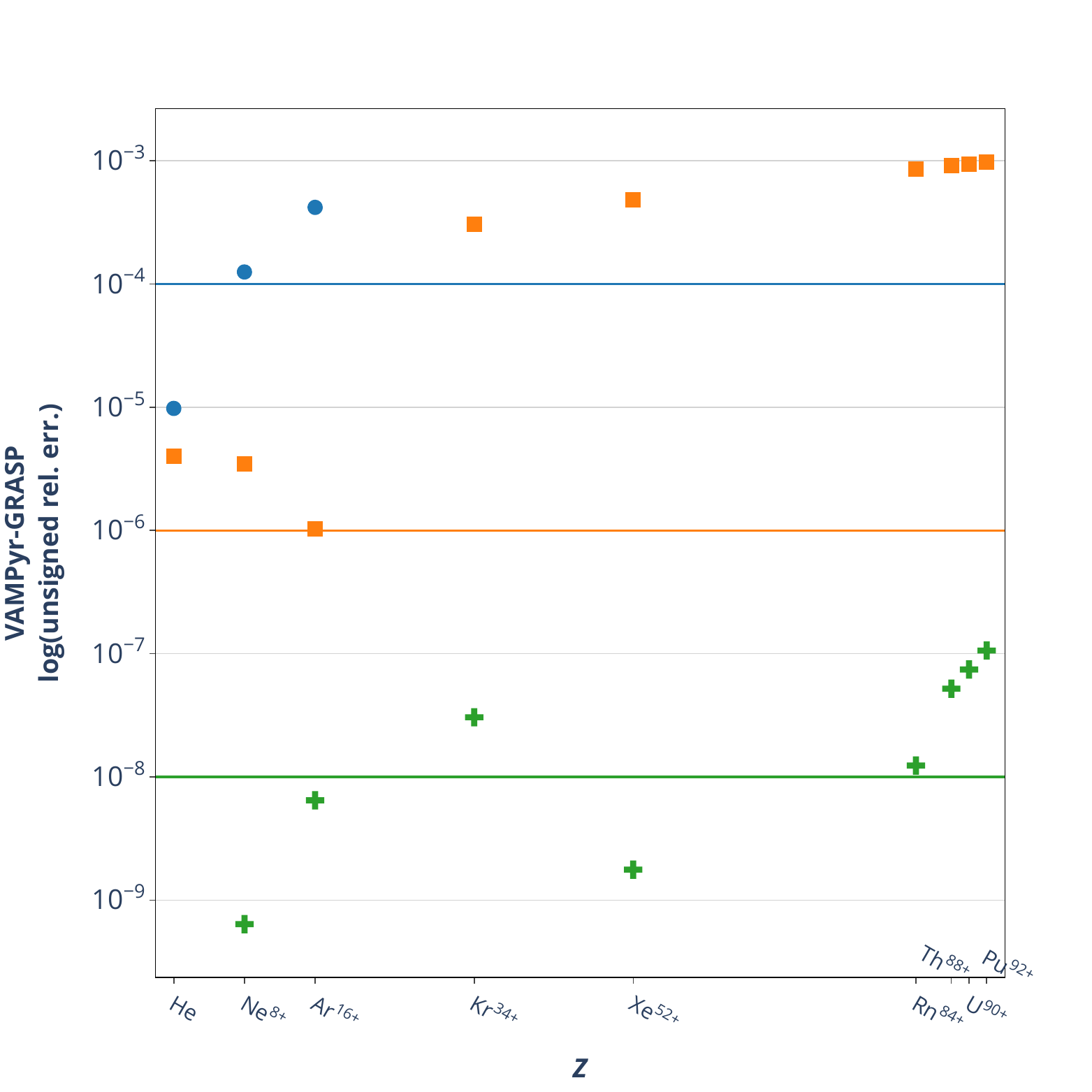}
    \caption{Logarithm of unsigned relative error between the
	Dirac-Coulomb-Hartree-Fock ground-state energy calculations from
	\vampyr{} and \grasp{}. All species
	are in the electronic configuration $1s^{2}$. The \vampyr{}
	calculations were done with different choices of Legendre polynomial order $k$ and
	tolerance $\epsilon$: blue circle, $k=6$, $\epsilon = 10^{-4}$; orange
	square, $k=8$, $\epsilon = 10^{-6}$; green cross, $k=10$, $\epsilon =
	10^{-8}$.  Both codes have used nuclear point charge model as
	described in Ref.~\citenum{Visscher_1997}.} 
    \label{fig:vampyr_vs_grasp-DCHF}
\end{figure}

Our multiwavelet implementation in \vampyr{} uses two parameters to tune
the precision of the calculation: the tolerance, $\varepsilon$, and the
polynomial order, $k$. Furthermore, both point-charge and Fermi models are
available for the nuclei.
In order to validate our \ac{DCHF} implementation and reassess the impact of
the nuclear model, we performed \ac{DCHF} calculations with a point-charge
model and increasingly tighter precision settings. We report the comparison of
our results with \grasp{} in Figure~\ref{fig:vampyr_vs_grasp-DCHF}. 
The relative errors obtained at looser precision settings, as shown in
Fig.~\ref{fig:vampyr_vs_grasp-DCHF}, are not consistent with the user-requested
$\epsilon$ for heavy elements.
The desired precision is user-selected through the settings for $\epsilon$ and
$k$ and should, in principle, be achieved irrespective of the nuclear model.
However, our results show that a point-charge nuclear model can
reproduce grid-based results from \grasp{} only when a very tight
tolerance is chosen, see Fig.~\ref{fig:vampyr_vs_grasp-DCHF} and Table 1 in
\ac{SI}. 
At the opposite end, SCF convergence could not be achieved for $k=6$, $\epsilon =
10^{-4}$ for \ce{Kr^{34+}} and heavier elements. 

One possible explanation is the choice of point-like nuclear potential, which is nonphysical and not suitable for fully relativistic calculations, but only for nonrelativistic ones.
Thus, calculations with a point nucleus require a significant tighter \textit{tolerance} and consequently a higher polynomial order to achieve the same precision of grid-based results from \grasp{}.

After assessing the validity of our method for the \ac{DCHF} equation, we
developed the Gaunt and Gauge two-electron terms in the Breit Hamiltonian as a
perturbative correction, as done in \emph{GRASP}.
The Gaunt term contains the vector operator $\Vec{\alpha}$: it is a Cartesian
vector of $4\times 4$ matrices whose antidiagonal blocks are the Pauli matrices
for the corresponding Cartesian direction. 
As we have previously mentioned in the \textit{Introduction}, it can be seen as
the curl of a spinorbital in the classical limit \cite{Saue_2011}. 
$\Vec{\alpha}$ acting on a 4-component orbital mixes its components to give the
the current density generated by the rotation of the spin around its axis
\cite{Saue_2011}.

We first compared \ac{DCHF} results from \dirac{} with those
obtained with \vampyr{} at high precision (\emph{i.e.} $k=10$, $\epsilon
= 10^{-8}$), see Table 2 in \ac{SI}.
These results confirm and extend to the \emph{full} 4-component regime the
observations of Jensen \textit{et al.}: \acp{MW} can attain higher precision than
large Gaussian atomic basis sets.\cite{Jensen_2017}

\begin{figure}[!ht]
    \centering
        \begin{subfigure}[t]{0.45\textwidth}
        \centering
        \includegraphics[width=\textwidth]{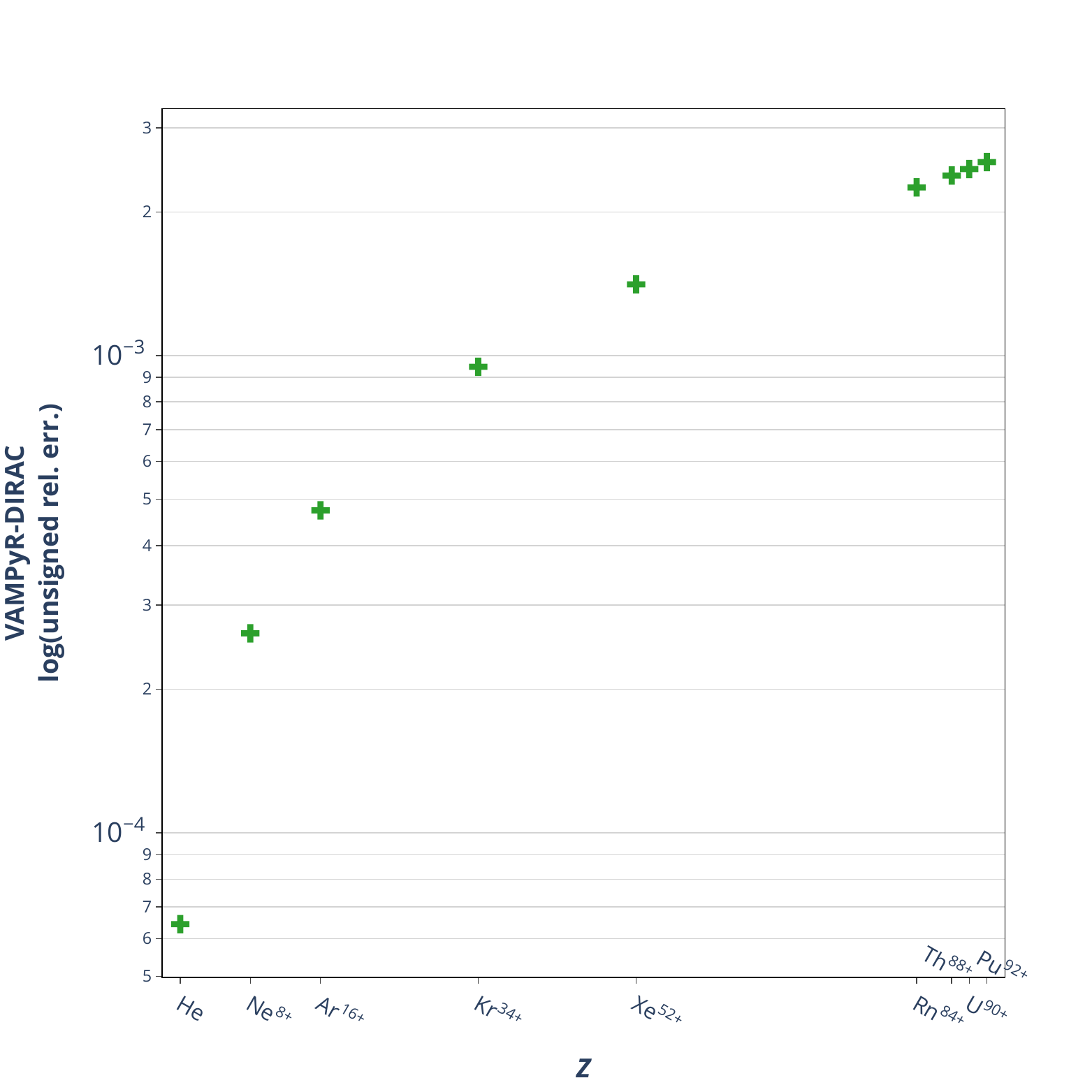}
        \caption{}
        \label{fig:spinorbit}
    \end{subfigure}
    \begin{subfigure}[t]{0.45\textwidth}
        \centering
        \includegraphics[width=\textwidth]{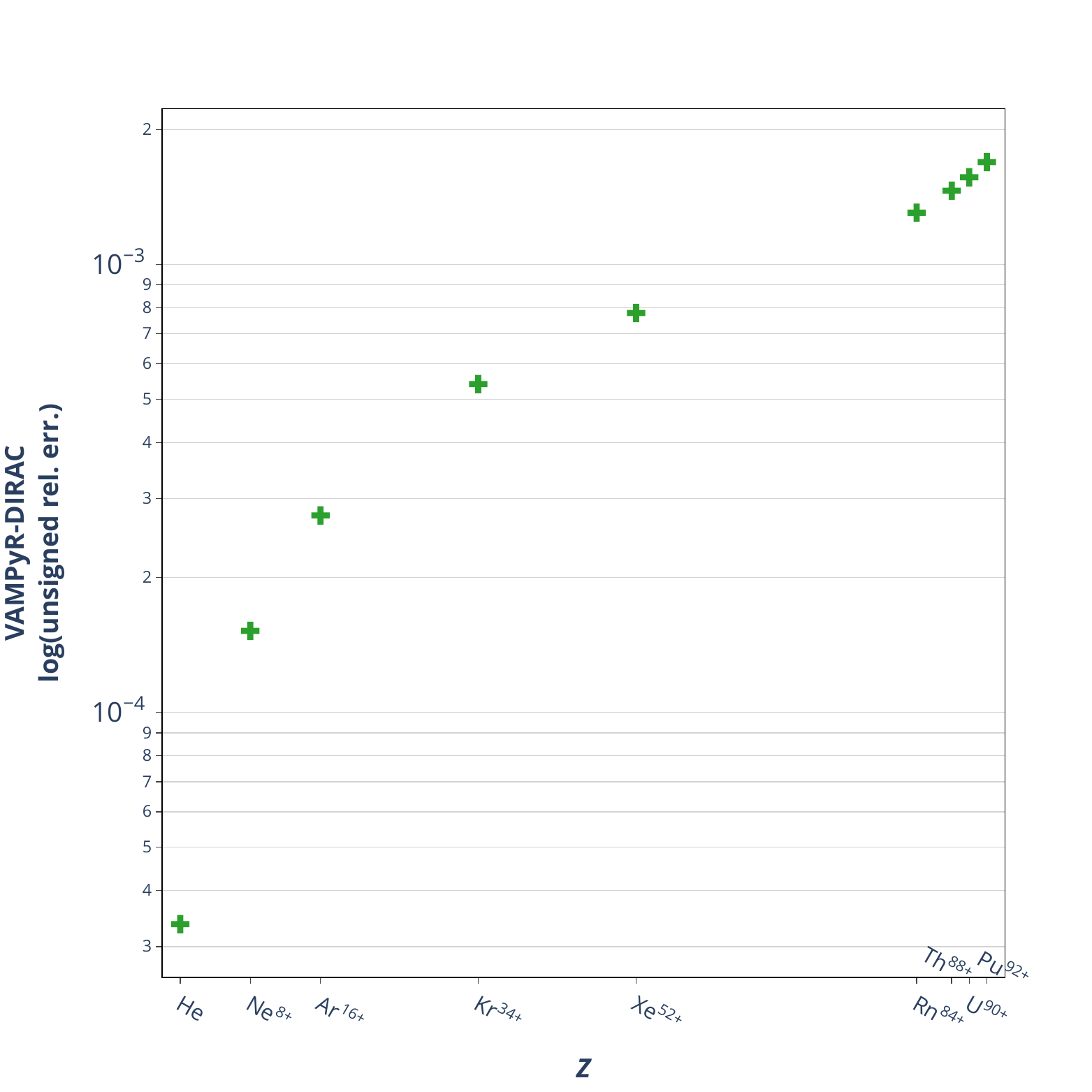}
        \caption{}
        \label{fig:gaunt}
    \end{subfigure}

    \caption{Comparison between the spinorbit energies (a) and Gaunt terms (b)
	coming from \vampyr{} and \dirac{} for selected systems in electronic configuration $1s^{2}$. The $y$ axis shows the
	logarithm of the unsigned relative difference between \vampyr{}
	and \dirac{} results. The \vampyr{} calculations were done
	with Legendre polynomial order $k=10$ and tolerance $\epsilon =
	10^{-8}$.  All codes have used a nuclear point charge model as
	described in Ref.~\citenum{Visscher_1997}.} 
    \label{fig:vampyr_vs_dirac}
\end{figure}

Thereafter, we compared our perturbative Gaunt correction, implemented in
\vampyr{}, with the variational implementation available in the
\dirac{} code, see Figure~\ref{fig:vampyr_vs_dirac}.
The inclusion of the Gaunt term in the variational self-consistent field
procedure is not expected to significantly affect the ground state as
previously shown\cite{Thierfelder_2010} and both results can be compared, see
Figure~\ref{fig:vampyr_vs_dirac}. 
In fact, the logarithm of the unsigned relative errors for the spinorbit energies,
see Fig.~\ref{fig:spinorbit}, and the Gaunt terms, see Fig.~\ref{fig:gaunt},
between \vampyr{} and \dirac{} have the same order of magnitude.

The perturbative Gauge correction only involves the inverse interelectronic distance kernel, as shown in by Eqs.~\eqref{eq:gauge-4-energy-1term}-~\eqref{eq:gauge-4-energy-6term},
from which it is evident how the magnetic energy term arises as half of
the Gaunt term, since both the direct and exchange Gauge contributions (third
and sixth terms) contain half of the Gaunt term.

\begin{figure}[!ht]
    \centering
    \includegraphics[width=\textwidth]{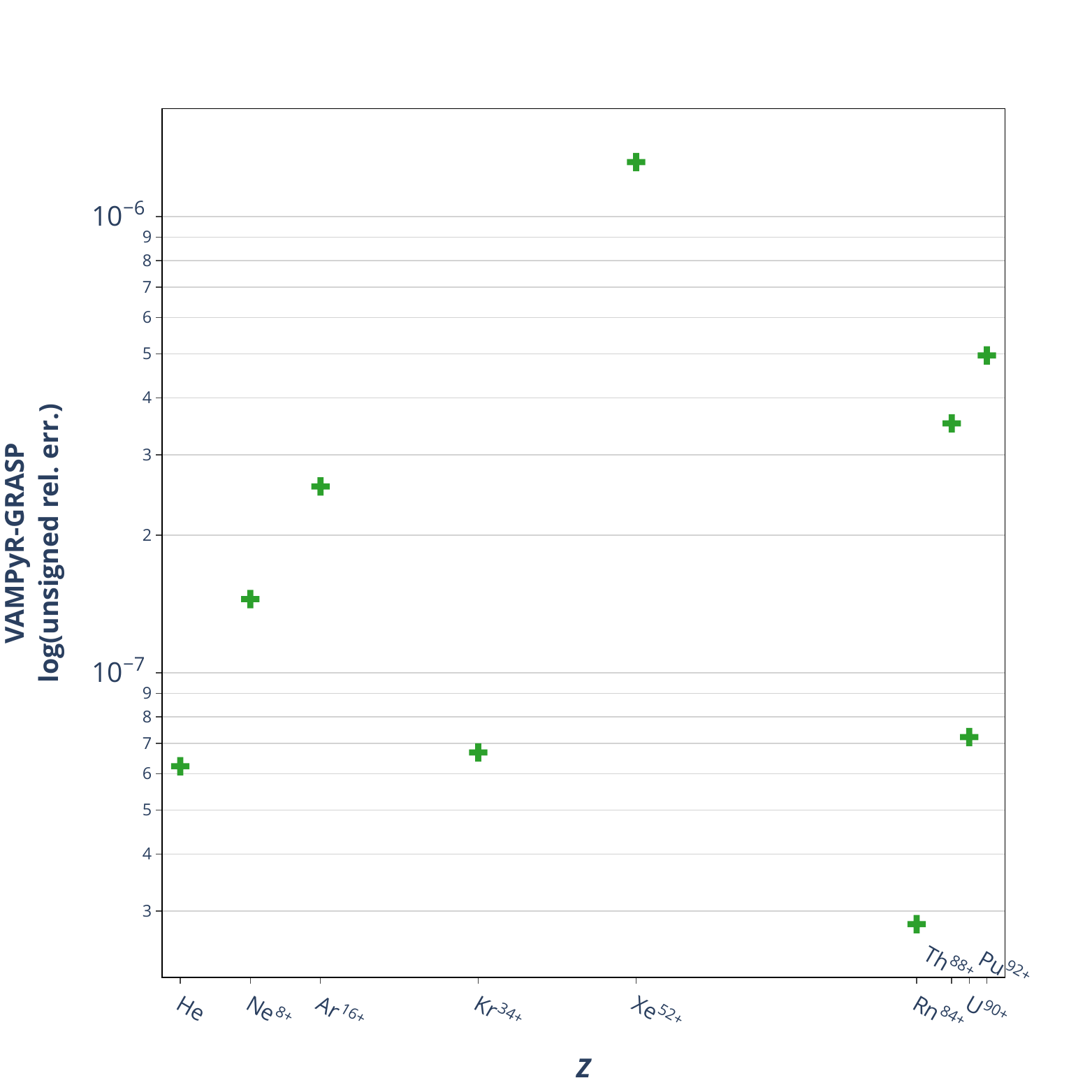}
    \caption{Comparison between the Breit perturbative corrections computed
	\vampyr{} and \grasp{} for noble gases and actinides in
	electronic configuration $1s^{2}$. The $y$ axis shows the logarithm of
	the unsigned relative difference between \vampyr{} and
	\grasp{} results. The \vampyr{} calculations were done with
	Legendre polynomial order $k=10$ and tolerance $\epsilon = 10^{-8}$.
	All codes have used a nuclear point charge model as described in
	Ref.~\citenum{Visscher_1997}.} 
    \label{fig:vampyr_vs_grasp-Breit}
\end{figure}

For the specific case of $1s^{2}$ systems, the terms involving a gradient in the Gauge energy (i.e., 1st Eq.~\eqref{eq:gauge-4-energy-1term}, 2nd Eq.~\eqref{eq:gauge-4-energy-2term}, 4th Eq.~\eqref{eq:gauge-4-energy-4term} and 5th Eq.~\eqref{eq:gauge-4-energy-5term}) are either zero or cancel each other out, up to the chosen
numerical precision $\varepsilon$. 
Thus, the ratio between the Gauge term $(E^{Gauge})$ and the magnetic
interaction energy, which corresponds to half of the Gaunt term $(E^{Mag} = \frac{1}{2}
E^{Gaunt})$, should be one (\emph{i.e.}, identical Magnetic and Gauge terms). 
This was verified comparing the Breit energy corrections from \vampyr{}
and \grasp{} results, see Figure~\ref{fig:vampyr_vs_grasp-Breit} and
Table 5 in the \ac{SI}.

The $E^{Gauge}/E^{Mag}$ ratio was calculated previously using Gaussian atomic
orbital basis sets for several atoms from $Z=9$ to $Z=79$ \cite{Sun_2022}.
It was shown to range between 0.90 (Fluorine) and 0.80 for $Z > 56$, converging asymptotically.
In Table 5 of the \ac{SI}, where we have considered $1s^2$ systems exclusively, we have obtained a unitary ratio between Gauge and magnetic term.
Furthermore, the magnitude of the Gauge term from our results in Table 5 in \ac{SI} confirms
what was previously found by Halbert \textit{et al.} \cite{Halbert_2021} that
in core-electron spectroscopy the Gauge term remains quite significant for the
$K$ and $L$ edges, and it must be accounted for, especially for $1s$ to $2s$
transitions \cite{Boudjemia_2019}.

\section{Conclusions}

We have shown that the 4-component Dirac-Coulomb-Hartree-Fock equations can be
solved self-consistently with a fully adaptive \ac{MW} basis irrespective of the chosen nuclear model, as required with Gaussian basis sets.\cite{Ishikawa_1987,Visser_1987}

The use of \ac{MRA} with a \ac{MW} basis to solve the KS-DFT equations allows to separate model errors from discretization (\emph{i.e.} basis set) errors, with the latter precisely quantifiable. Thus, the use of a MW basis provides fundamental insight to understand the range of applicability of KS-DFT with localized basis sets. This issue is especially relevant for 4-component relativistic calculations on heavy elements where the description of the core electrons is challenging due the nature of the Dirac equation combined with the extremely high nuclear charge and a reduced availability of \ac{GTO} bases.

We have shown that the \ac{DCHF} ground state combined with the Breit Hamiltonian as a
perturbative correction can reproduce grid-based calculations performed with
\grasp{}. Albeit not performed in this work, the fully variational inclusion of
the Gaunt and Gauge terms can be obtained by making use of the corresponding
operator expressions (Equations~\eqref{eq:Gaunt-direct} and
\eqref{eq:Gaunt-exchange} and Equations~\eqref{eq:Gauge-direct} and
\eqref{eq:Gauge-exchange} for Gaunt and Gauge, respectively). This has not been
done for the current work both to simplify the comparison with the \grasp{}
code and because of excessive memory demands of the current pilot
implementation. The latter is indeed the main challenge for future extensions
to general molecular systems where the simplifications that enabled our results
(time-reversal symmetry, spherical symmetry of the $1s$ orbital) will no longer
hold. Work is in progress in our group to overcome these hurdles.

The unitary $E^{Gauge}/E^{Mag}$ ratio for $s$-orbitals explains how neither
Gaunt nor Gauge terms can be neglected for core-electron spectroscopy and
explains the importance of considering both these terms when x-ray photoelectron
spectra are calculated to fit the experimental ones
\cite{Halbert_2021,Boudjemia_2019,Oura_2019}.
Our results confirm the validity of the \ac{MW} approach for future
development of core-electron spectroscopy to resolve the structures of oxides of
$f$-elements and other strongly correlated systems.




\begin{acknowledgement}
We would like to thank Prof.~Trond Saue from the CNRS/Université de
Toulouse, France, Dr.~Jon Grumer from Uppsala University, Sweden, and
Dr.~Michal Repisky from UiT, The Arctic University of Norway for useful
discussions. We acknowledge support from the Research Council of Norway through
its Centres of Excellence scheme (262695), through the FRIPRO grant ReMRChem
(324590),  and from NOTUR -- The Norwegian Metacenter for Computational Science
through grant of computer time (nn4654k).
\end{acknowledgement}

\begin{suppinfo}
All data generated or analyzed during this study are included in the graph showed in this article and tables showed in supporting information.
\end{suppinfo}

\bibliography{sample}
\end{document}